\documentclass[preprint,showpacs,preprintnumbers,amsmath,amssymb,nofootinbib]{revtex4}
\usepackage{epsfig}
\usepackage{epstopdf}
\usepackage{bbm}
\usepackage{color}
\newcommand{\be}{\begin{equation}}
\newcommand{\ee}{\end{equation}}
\newcommand{\bea}{\begin{eqnarray}}
\newcommand{\eea}{\end{eqnarray}}

\newcommand{\bino}{{\tilde B}}
\newcommand{\wino}{{\tilde W}}
\newcommand{\higgsino}{{\tilde H}}
\newcommand{\slepton}{{\tilde l}}

\newcommand{\TeV}{{~\rm TeV}}
\newcommand{\Arg}{{\rm Arg}}
\newcommand{\gm}{{$g-2$~}}
\def\simlt{\mathrel{\raise.3ex\hbox{$<$\kern-.75em\lower1ex\hbox{$\sim$}}}}
\def\simgt{\mathrel{\raise.3ex\hbox{$>$\kern-.75em\lower1ex\hbox{$\sim$}}}}

\usepackage{scalerel}[2014/03/10]
\usepackage{stackengine}

\begin{document}
%\pagestyle{plain}
%\renewcommand{\thefootnote}{\fnsymbol{footnote}}
%preprint{CUMQ/HEP xxx}
%\preprint{}
%
%\preprint{IPMU18-XXXX}

\title{ Muon \gm and CP violation in MSSM}

\author{Chengcheng Han$^1$ \\~ \vspace*{-0.4cm}}
\affiliation{
$^1$ School of Physics, Sun Yat-Sen University, Guangzhou 510275, China
  \vspace*{1.5cm}
}

\begin{abstract}

%We study the CP violation effect on the parameter space to explain the muon \gm in minimal supersymmetric standard model. of the minimal supersymmetric standard model(MSSM) 

We study the constraints of the CP violation in the muon \gm preferred region of the minimal supersymmetric standard model assuming a universal slepton masses within first two generations. We present two particular scenarios where the \gm anomaly is predicted within 2 $\sigma$ level mainly through the chargino loop or the bino loop. We found that for both cases the electron EDM experiment already highly constrained the CP phase of the parameters: either the Arg[$\mu M_1]$ or Arg[$\mu M_2]$ should be smaller than $\mathcal O$(2-3)$\times10^{-5}$. If the muon \gm anomaly is explained by the MSSM, a particular SUSY breaking mechanism is needed to guarantee the small CP phase of SUSY parameters. Otherwise, a tuning of $\mathcal O(10^{-5})$ is needed to cancel the phase in a general CP violated SUSY model.

%the constraints from the electron edm on
\end{abstract}
%\pacs{}

\maketitle

%%%%%%%%%%%%%%%%%%%%%%%%%%%%%%%%%%%%%%%%%%%%%%%%%%%%%%%%%%%%%%
%\newpage
%\tableofcontents
\section{Introduction}

The minimal supersymmetric standard model(MSSM) as the minimal extension of the standard model in the framework of supersymmetry, remains one of the well-motivated models beyond the standard model. It not only alleviates the Higgs mass hierarchy problem but also provides a dark matter candidate as well as predicts the unification of the gauge couplings at a high energy scale. Therefore it is also one of the most important targets to search for at high-energy physics experiments. However, the large hadron collider(LHC) already set a very strong limit on the masses of SUSY particles. For example, the gluino and first two generation squarks should be heavier than 2 TeV , and the electroweak particles should beyond a few hundred GeV \cite{Aad:2020nyj, Aad:2020sgw, Aad:2019vvi, Aad:2019qnd} . 

Since there are still no signatures of SUSY particles in direct search experiments. It is intriguing to consider that whether the precision measurement could provide any indirect signature of the SUSY particles. For example, experiments to measure the lepton flavor-related process such as $\mu \rightarrow e \gamma$, muon $g-2$, electron EDM, etc, can reach very high precision and may provide a good signature of high scale physics. Interestingly, the Fermilab just announced the new measurement of the muon anomalous magnetic momentum~\cite{fermilab} and confirm previous search results~\cite{Bennett:2006fi}, indicating a large deviation of SM prediction at 4.2$\sigma$~\cite{Aoyama:2020ynm} after combining:
\begin{eqnarray}
\Delta a_\mu = (25.1 \pm 5.9) \times 10^{-10}
\end{eqnarray}
It would be very interesting if such deviation can be explained in the framework of supersymmetry. Indeed there are already numerous studies on these \cite{Moroi:1995yh, Endo:2017zrj, Cox:2018vsv, Cox:2018qyi,  gm2} and it shows that if smuon masses are less than TeV  such anomaly can be easily explained.
 
On the other hand, to avoid the large FCNC process in the quark sector, a flavor-blinded mediation of SUSY breaking models are usually preferred. For example, in the gauge mediation models, the gaugino mediation models, or the anomaly mediation models, a universal squark masses is usually predicted at a high energy scale. However, such kinds of models also predict a universal masses for the slepton sector, resulting in a close mass parameters for the first two generation sleptons at low energy scale\footnote{The third generation sfermions could be much different at low energy scale due to the large RGE effect from the Yukawa sector~\cite{Cox:2018vsv}.}. If the SUSY parameter is generally CP violated, it would unavoidably induce large CP violation in the electron sector as well. The measurement of electron EDM already highly constrained the CP violation effect in the electron sector~\cite{Andreev:2018ayy}
\begin{eqnarray}
|d_e| < 1.1\times 10^{-29} 
\end{eqnarray}
This fact intriguing us that the muon \gm and electron EDM might be correlated and the CP phase of the SUSY parameters should be constrained.

In this paper, we study the constraints of electron EDM on the CP violation phase of the SUSY parameter space where the muon \gm anomaly can be explained. This article is organized as follows, in Sec.~\ref{gm2} we briefly discuss the physics related to the muon \gm and electron EDM. In Sec.~\ref{numerical} we present our strategy of numerical calculation and the result. We draw our conclusion in Sec. \ref{conclusion}.

\section{Muon \gm and electron EDM in MSSM}
\label{gm2}

In the MSSM, if the selectron and smuon share same mass parameters, the similar diagram contributing the muon \gm would also contribute to the electron EDM if the parameters are generally CP violated. In the following, we discuss the physics related to muon \gm and electron EDM.
\subsection{Muon \gm }

In MSSM, there are five dominant contributions to the muon \gm \cite{Moroi:1995yh, Endo:2017zrj} which can be divided into the neutralino-slepton loop and chargino-sneutrino loop. In the neutralino-slepton scenario, there are four dominant contributions: the $\bino-\higgsino-\slepton_R$, $\bino-\higgsino-\slepton_L$, $\bino-\slepton_L-\slepton_R$ and $\wino-\higgsino-\slepton_L$ loops . The first two cases require light sleptons as well as light $\bino$ and $\higgsino$ which is highly constrained by both LHC searches and the dark matter direct searches. However, in the $\bino-\slepton_R-\slepton_L$ scenario, due to the enhancement of the $\mu$-term, the slepton mass could be much heavier, providing an interesting possibility\footnote{There are also cases where different scenarios mixed, in this paper, we only consider one of the particular scenarios to be dominant.}. The corresponding mass parameters in this scenario are $\mu$, $M_1$, and slepton masses. In the case of the $\wino-\higgsino-\slepton_L$ loop, it also associates with the chargino-sneutrino loop. The parameters related to this scenario are $\mu$, $M_2$, and slepton masses. In this papers, we mainly focus on these two scenarios and they are denoted as bino loop and chargino loop respectively, which can be summarized as \cite{Moroi:1995yh, Endo:2017zrj}:

\begin{eqnarray}
a_\mu({ \bino-\slepton_L-\slepton_R}) \simeq \frac{\alpha_Y}{4\pi} \frac{m_\mu^2 M_1 \mu}{m^2_{\tilde \mu_L} m^2_{\tilde \mu_R}} \tan\beta f_N \left(   \frac{m^2_{\tilde \mu_L}}{M_1^2} , \frac{m^2_{\tilde \mu_R}}{M_1^2} \right)
\end{eqnarray}

\begin{eqnarray}
a_\mu({\wino-\higgsino-\slepton_L}) &\simeq& -\frac{\alpha_2}{8\pi} \frac{m_\mu^2 }{M_2 \mu} \tan\beta f_N \left(   \frac{M_2^2}{m^2_{\tilde \mu_L}} , \frac{\mu^2}{m^2_{\tilde \mu_L}} \right)   \nonumber \\
a_\mu({ \wino-\higgsino-\tilde \nu}) &\simeq&  \frac{\alpha_2}{4\pi} \frac{m_\mu^2 }{M_2 \mu} \tan\beta f_C \left(   \frac{M_2^2}{m^2_{\tilde \nu}} , \frac{\mu^2}{m^2_{\tilde \nu}} \right) 
\end{eqnarray}
where

\begin{eqnarray}
f_N(x,y) &=& xy \left[  \frac{-3+ x + y + x y}{(x-1)^2 (y-1)^2} + \frac{2 x \ln x}{(x-y) (x-1)^3}  -\frac{2 y \ln y}{(x-y) (y-1)^3}  \right]   \\
f_C(x,y) &=& xy \left[  \frac{5-3( x + y) + x y}{(x-1)^2 (y-1)^2} - \frac{2  \ln x}{(x-y) (x-1)^3}  +\frac{2 \ln y}{(x-y) (y-1)^3}  \right] 
\end{eqnarray}

\subsection{Electron EDM}

The operator related to the electron EDM is:
\begin{eqnarray}
\mathcal L = -\frac{i}{2} d_f \bar \psi \sigma_{\mu \nu} \gamma_5 \psi F^{\mu\nu}
\end{eqnarray}
The EDM of a particle with spinor $\psi_i$ and a scalar $\phi_k$ with the interaction contains CP violation can be given~\cite{Ibrahim:2007fb, Cesarotti:2018huy}
\begin{eqnarray}
\mathcal L &=& L_{ik} \bar \psi_f P_L \psi_i \phi_k + R_{ik} \bar \psi_f P_R \psi_i \phi_k + H.c.   \\
d_f &=&\frac{m_i}{16\pi^2 m_k^2} Im (L_{ik} R^*_{ik}) \left[Q_i A \left(\frac{m_i^2}{m_k^2}\right) + Q_k B\left (\frac{m_i^2}{m_k^2} \right) \right ]
\end{eqnarray}
where
\begin{eqnarray}
A(r) &=& \frac{1}{2(1-r)^2}(3-r+ \frac{2\ln r}{1-r})  \\
B(r) &=& \frac{1}{2(1-r)^2}(1+r+ \frac{2\ln r}{1-r})  
\end{eqnarray}

The $\psi_i$ here could be the chargino or neutralino and the $\phi$ here could be slepton or sneutrino. The interactions can be easily calculated after rotating the charginos and neutralinos as well as the sleptons into the mass basis.

\section{Numerical calculation and results}
\label{numerical}
In MSSM, the related parameter could have a phase are $\mu, A_e, A_\mu, M_1, M_2$. However, not all the phase are independent parameters. The physics only depends on the relative phase of them. To simplify our analysis,  we take  $A_e=A_\mu=0$, then there are only two independent phase $\Arg[\mu M_1]$ and $\Arg[\mu M_2]$ which could affect our result. In our scan, we assume $\tilde \mu_L = \tilde \mu_R = \tilde e_L= \tilde e_R =m_0$. Except for the parameters $M_2$, $M_1$, $\mu$ and $m_0$, all the other mass parameters are set to be 5 TeV and the corresponding phases are set to be 0. We also set $\tan\beta=50$ to accommodate the muon \gm.

To factor out the chargino loop and bino loop separately, we divide our scan into two categories. For the bino loop scenario, the related parameters are set as follows:
\begin{eqnarray}
&& M_2 = 0.5 \TeV, \tan \beta= 50, M_1= m_0 \nonumber \\
&& 0.5 \TeV < m_0 <  1.3 \TeV, 3 \TeV < \mu < 10 \TeV \nonumber \\
&& \Arg[\mu] =\Arg[M_2] =0, ~~10^{-7} <\Arg[M_1] < 10^{-1}.
\end{eqnarray}
The minimum of the $M_2=0.5 \TeV$ is due to the limit from LHC search for displaced chargino searches~\cite{Aaboud:2017mpt}. $\mu$ is larger than 3 TeV to accommodate the muon \gm for the bino loop, at the same time the chargino contribution can be diminished. Therefore the parameter space to explain muon \gm mainly from the bino loop. The scan of phase on $M_1$ is to satisfy the constraint of electron EDM on $\Arg[\mu M_1]$.

For the chargino loop scenario, we set
\begin{eqnarray}
&& \mu = 0.4 \TeV, \tan \beta= 50, M_1= 3 \TeV \nonumber \\
&& 0.4 \TeV < m_0 <  1.3 \TeV, 0.4 \TeV < M_2 < 2 \TeV \nonumber \\
&& \Arg[\mu] =\Arg[M_1] =0, ~~ 10^{-7} <\Arg[M_2] < 10^{-1}.
\end{eqnarray}
The lower limit on the $\mu$ is due to the LHC sleptons searches~\cite{Aad:2019vnb}. The updated ATLAS slepton search set a very strong limit on the slepton pair production for an LSP mass less than 400 GeV. The $M_1$ is set to be 3 TeV to reduce the bino loop contribution.

In our calculation, the mass spectrum and the muon \gm and electron EDM are calculated by {\bf CPsuperH 2.3} \cite{Lee:2003nta, Lee:2007gn, Cheung:2009fc}. The dark matter relic density and dark matter nucleon interaction are calculated by {\bf MicrOMEGAs 5.2} \cite{Belanger:2006is}.

\begin{figure}[htbp]
 \centering
 \includegraphics[width=0.45\textwidth]{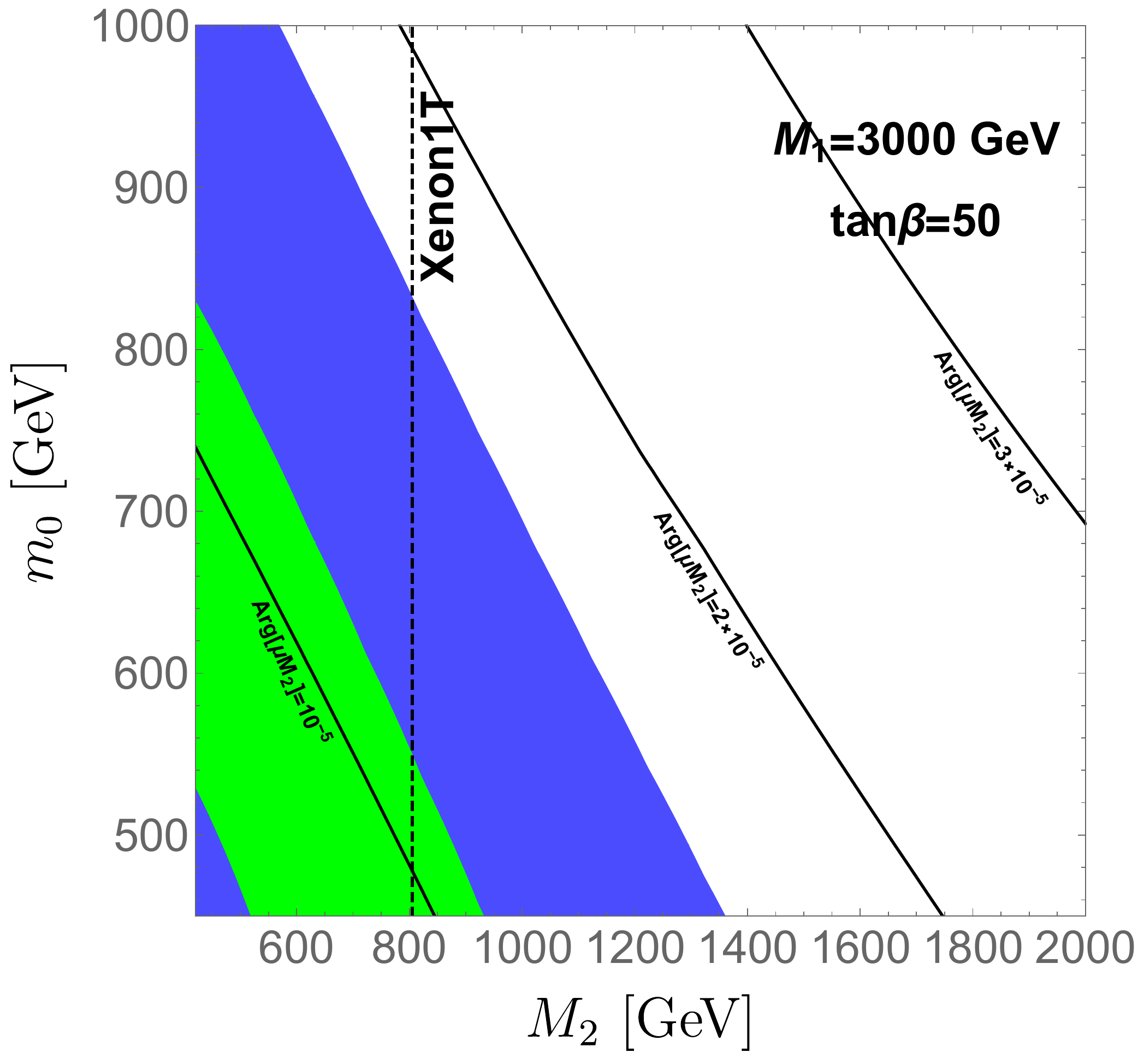} 
 \includegraphics[width=0.45\textwidth]{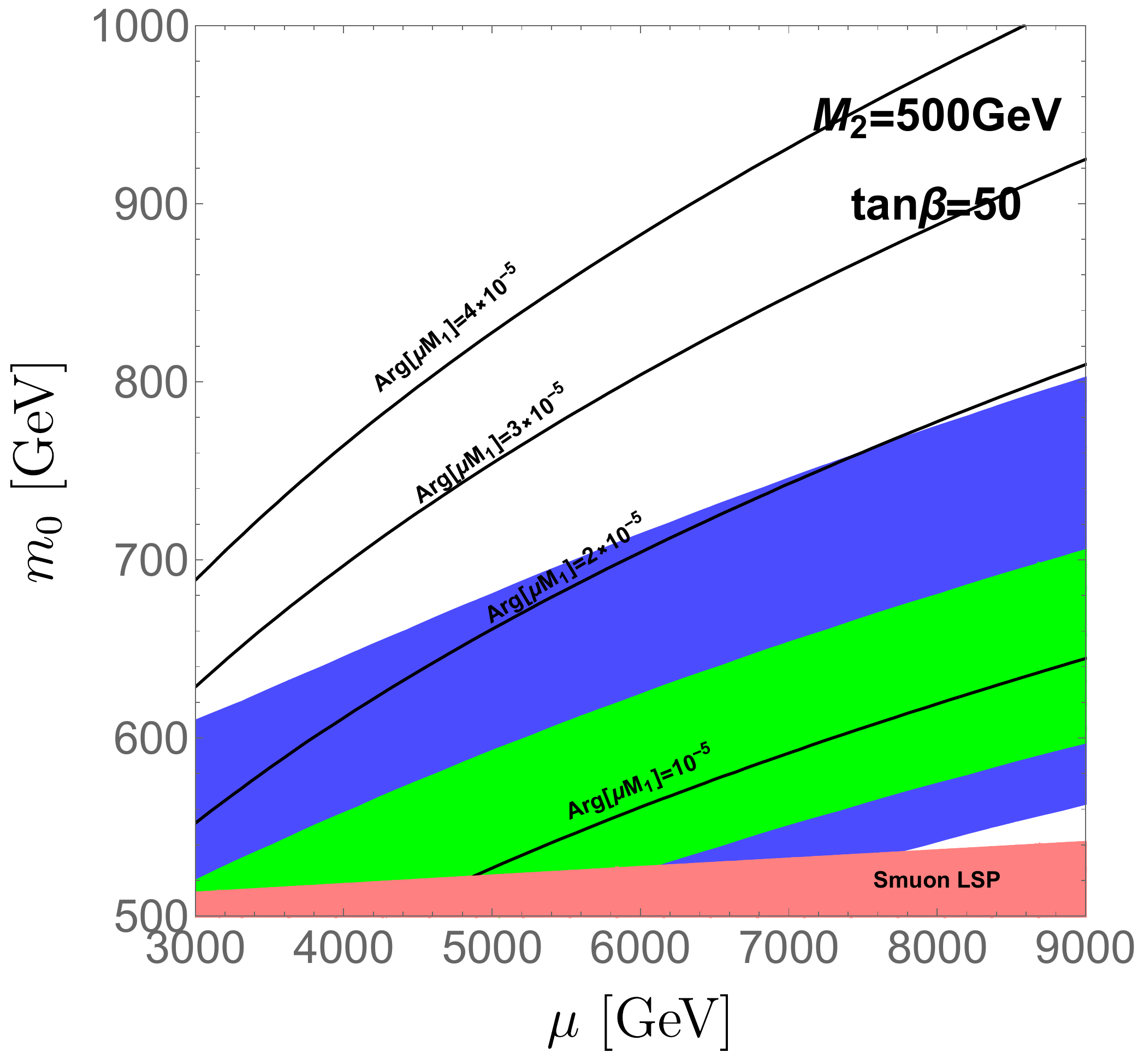} 
 \caption{The parameter space for two scenarios: chargino  loop contribution(left panel) and bino loop contribution(right panel). The real lines are the constraints from the electron EDM.}
 \label{fig:higgsino}
\end{figure}

Our final results are shown in Fig.~\ref{fig:higgsino}. On the left panel we show the parameter space to explain muon \gm for the chargino loop.  We can see that the current Xenon1T search already excluded part of $1\sigma$ preferred region, leaving a small parameter region to explain the \gm at 1$\sigma$ level. Since the higgsino dark matter is under abundance, a reduction factor is already included in the dark matter direct detection. In the plot we already set a limit on the electron EDM everywhere. Nevertheless, the phase of $\Arg[\mu M_2]$ should smaller than $2\times 10^{-5}$ if the muon \gm anomaly is explained by the chargino loop.

On the right panel, we show the parameter space for the \gm and the constraints on the phase of $\Arg[\mu M_1] $  in the case of bino loop contribution.  It tells that the phase of $\Arg[\mu M_1] $ should also smaller than $3\times 10^{-5}$ if the \gm is explained primarily by the bino loop. We note this limit is rather robust with $\tan\beta$ because both value of \gm and electron EDM proportional to $\tan \beta$.

\section{Summary and Conclusions}
\label{conclusion}
We study the constraints of the CP violation in the SUSY parameter space where the muon \gm anomaly can be explained. We present two particular scenarios where the \gm anomaly is predicted within 2 $\sigma$ level mainly through the chargino loop or the bino loop. We found that for both cases the electron EDM experiment already highly constrained the CP phase of the parameters: either the Arg[$\mu M_1]$ or Arg[$\mu M_2]$ should be smaller than $\mathcal O$(2-3)$\times10^{-5}$. If the muon \gm anomaly is explained by the MSSM, a particular SUSY breaking mechanism is needed to guarantee the small CP phase of SUSY parameters\footnote{One of such models is shown in \cite{Evans:2020vil}.}. Otherwise, a tuning of $\mathcal O(10^{-5})$ is needed to cancel the phase in a general CP violated SUSY model.

\section*{Acknowledgements}
C. Han thank Song Li, Peiwen Wu and Jin Min Yang for helpful discussions and useful suggestions. C.H. acknowledges support from the Sun Yat-Sen University Science Foundation.


\begin{thebibliography}{99}

\bibitem{Aad:2020nyj}
G.~Aad \textit{et al.} [ATLAS],
%``Search for new phenomena in final states with large jet multiplicities and missing transverse momentum using $ \sqrt{s} $ = 13 TeV proton-proton collisions recorded by ATLAS in Run 2 of the LHC,''
JHEP \textbf{10} (2020), 062
%doi:10.1007/JHEP10(2020)062
[arXiv:2008.06032 [hep-ex]].
%3 citations counted in INSPIRE as of 11 Jan 2021

%\cite{Aad:2020sgw}
\bibitem{Aad:2020sgw}
G.~Aad \textit{et al.} [ATLAS],
%``Search for a scalar partner of the top quark in the all-hadronic $t{\bar{t}}$ plus missing transverse momentum final state at $\sqrt{s}=13$ TeV with the ATLAS detector,''
Eur. Phys. J. C \textbf{80} (2020) no.8, 737
%doi:10.1140/epjc/s10052-020-8102-8
[arXiv:2004.14060 [hep-ex]].
%17 citations counted in INSPIRE as of 11 Jan 2021

%\cite{Aad:2019vvi}
\bibitem{Aad:2019vvi}
G.~Aad \textit{et al.} [ATLAS],
%``Search for chargino-neutralino production with mass splittings near the electroweak scale in three-lepton final states in $\sqrt {s}$=13  TeV $pp$ collisions with the ATLAS detector,''
Phys. Rev. D \textbf{101} (2020) no.7, 072001
%doi:10.1103/PhysRevD.101.072001
[arXiv:1912.08479 [hep-ex]].
%23 citations counted in INSPIRE as of 11 Jan 2021

%\cite{Aad:2019qnd}
\bibitem{Aad:2019qnd}
G.~Aad \textit{et al.} [ATLAS],
%``Searches for electroweak production of supersymmetric particles with compressed mass spectra in $\sqrt{s}=$ 13 TeV $pp$ collisions with the ATLAS detector,''
Phys. Rev. D \textbf{101} (2020) no.5, 052005
%doi:10.1103/PhysRevD.101.052005
[arXiv:1911.12606 [hep-ex]].
%44 citations counted in INSPIRE as of 11 Jan 2021



%\cite{Aad:2019qnd}
\bibitem{fermilab}
Muon g ? 2 Collaboration, Physical Review Letters 126, 141801 (2021)


%\cite{Bennett:2006fi}
\bibitem{Bennett:2006fi}
G.~W.~Bennett \textit{et al.} [Muon g-2],
%``Final Report of the Muon E821 Anomalous Magnetic Moment Measurement at BNL,''
Phys. Rev. D \textbf{73} (2006), 072003
%doi:10.1103/PhysRevD.73.072003
[arXiv:hep-ex/0602035 [hep-ex]].
%2339 citations counted in INSPIRE as of 04 Apr 2021

%\cite{Aoyama:2020ynm}
\bibitem{Aoyama:2020ynm}
T.~Aoyama, N.~Asmussen, M.~Benayoun, J.~Bijnens, T.~Blum, M.~Bruno, I.~Caprini, C.~M.~Carloni Calame, M.~C\`e and G.~Colangelo, \textit{et al.}
%``The anomalous magnetic moment of the muon in the Standard Model,''
Phys. Rept. \textbf{887} (2020), 1-166
doi:10.1016/j.physrep.2020.07.006
[arXiv:2006.04822 [hep-ph]].
%140 citations counted in INSPIRE as of 07 Apr 2021







%\cite{Moroi:1995yh}
\bibitem{Moroi:1995yh}
T.~Moroi,
%``The Muon anomalous magnetic dipole moment in the minimal supersymmetric standard model,''
Phys. Rev. D \textbf{53} (1996), 6565-6575
[erratum: Phys. Rev. D \textbf{56} (1997), 4424]
%doi:10.1103/PhysRevD.53.6565
[arXiv:hep-ph/9512396 [hep-ph]].
%549 citations counted in INSPIRE as of 04 Apr 2021

%\cite{Endo:2017zrj}
\bibitem{Endo:2017zrj}
M.~Endo, K.~Hamaguchi, S.~Iwamoto and K.~Yanagi,
%``Probing minimal SUSY scenarios in the light of muon $g-2$ and dark matter,''
JHEP \textbf{06} (2017), 031
%doi:10.1007/JHEP06(2017)031
[arXiv:1704.05287 [hep-ph]].
%20 citations counted in INSPIRE as of 04 Apr 2021

%\cite{Cox:2018vsv}
\bibitem{Cox:2018vsv}
P.~Cox, C.~Han, T.~T.~Yanagida and N.~Yokozaki,
%``Gaugino mediation scenarios for muon $g?2$ and dark matter,''
JHEP \textbf{08} (2019), 097
%doi:10.1007/JHEP08(2019)097
[arXiv:1811.12699 [hep-ph]];

%\cite{Cox:2018qyi}
\bibitem{Cox:2018qyi}
P.~Cox, C.~Han and T.~T.~Yanagida,
%``Muon $g-2$ and dark matter in the minimal supersymmetric standard model,''
Phys. Rev. D \textbf{98} (2018) no.5, 055015
%doi:10.1103/PhysRevD.98.055015
[arXiv:1805.02802 [hep-ph]];



\bibitem{gm2}
Here only part of related papers are included:


%\cite{Ajaib:2015yma}
M.~A.~Ajaib, B.~Dutta, T.~Ghosh, I.~Gogoladze and Q.~Shafi,
%``Neutralinos and sleptons at the LHC in light of muon $(g-2)_{\mu}$,''
Phys. Rev. D \textbf{92} (2015) no.7, 075033
%doi:10.1103/PhysRevD.92.075033
[arXiv:1505.05896 [hep-ph]];
%40 citations counted in INSPIRE as of 04 Apr 2021
%\cite{Kobakhidze:2016mdx}
A.~Kobakhidze, M.~Talia and L.~Wu,
%``Probing the MSSM explanation of the muon g-2 anomaly in dark matter experiments and at a 100 TeV $pp$ collider,''
Phys. Rev. D \textbf{95} (2017) no.5, 055023
%doi:10.1103/PhysRevD.95.055023
[arXiv:1608.03641 [hep-ph]];
%24 citations counted in INSPIRE as of 04 Apr 2021
%15 citations counted in INSPIRE as of 07 Apr 2021
%8 citations counted in INSPIRE as of 07 Apr 2021
%\cite{Han:2020exx}
C.~Han, M.~L.~L\'opez-Ib\'a\~nez, A.~Melis, \'O.~Vives, L.~Wu and J.~M.~Yang,
%``LFV and (g-2) in non-universal SUSY models with light higgsinos,''
JHEP \textbf{05} (2020), 102
%doi:10.1007/JHEP05(2020)102
[arXiv:2003.06187 [hep-ph]];
%3 citations counted in INSPIRE as of 07 Apr 2021
%\cite{Chakraborti:2021kkr}
M.~Chakraborti, S.~Heinemeyer and I.~Saha,
%``Improved $(g-2)_\mu$ Measurements and Wino/Higgsino Dark Matter,''
[arXiv:2103.13403 [hep-ph]];
%0 citations counted in INSPIRE as of 07 Apr 2021
M.~Abdughani, K.~I.~Hikasa, L.~Wu, J.~M.~Yang and J.~Zhao,
%``Testing electroweak SUSY for muon $g$ \ensuremath{-} 2 and dark matter at the LHC and beyond,''
JHEP \textbf{11} (2019), 095
%doi:10.1007/JHEP11(2019)095
[arXiv:1909.07792 [hep-ph]].
%10 citations counted in INSPIRE as of 07 Apr 2021







%\cite{Andreev:2018ayy}
\bibitem{Andreev:2018ayy}
V.~Andreev \textit{et al.} [ACME],
%``Improved limit on the electric dipole moment of the electron,''
Nature \textbf{562} (2018) no.7727, 355-360
%doi:10.1038/s41586-018-0599-8
%300 citations counted in INSPIRE as of 07 Apr 2021








%\cite{Ibrahim:2007fb}
\bibitem{Ibrahim:2007fb}
T.~Ibrahim and P.~Nath,
%``CP Violation From Standard Model to Strings,''
Rev. Mod. Phys. \textbf{80} (2008), 577-631
%doi:10.1103/RevModPhys.80.577
[arXiv:0705.2008 [hep-ph]].
%120 citations counted in INSPIRE as of 07 Apr 2021

%\cite{Cesarotti:2018huy}
\bibitem{Cesarotti:2018huy}
C.~Cesarotti, Q.~Lu, Y.~Nakai, A.~Parikh and M.~Reece,
%``Interpreting the Electron EDM Constraint,''
JHEP \textbf{05} (2019), 059
%doi:10.1007/JHEP05(2019)059
[arXiv:1810.07736 [hep-ph]].
%41 citations counted in INSPIRE as of 07 Apr 2021












%\cite{Aaboud:2017mpt}
\bibitem{Aaboud:2017mpt}
M.~Aaboud \textit{et al.} [ATLAS],
%``Search for long-lived charginos based on a disappearing-track signature in pp collisions at $ \sqrt{s}=13 $ TeV with the ATLAS detector,''
JHEP \textbf{06} (2018), 022
%doi:10.1007/JHEP06(2018)022
[arXiv:1712.02118 [hep-ex]].
%149 citations counted in INSPIRE as of 07 Apr 2021

%\cite{Aad:2019vnb}
\bibitem{Aad:2019vnb}
G.~Aad \textit{et al.} [ATLAS],
%``Search for electroweak production of charginos and sleptons decaying into final states with two leptons and missing transverse momentum in $\sqrt{s}=13$ TeV $pp$ collisions using the ATLAS detector,''
Eur. Phys. J. C \textbf{80} (2020) no.2, 123
%doi:10.1140/epjc/s10052-019-7594-6
[arXiv:1908.08215 [hep-ex]].
%74 citations counted in INSPIRE as of 07 Apr 2021





%\cite{Lee:2003nta}
\bibitem{Lee:2003nta}
J.~S.~Lee, A.~Pilaftsis, M.~Carena, S.~Y.~Choi, M.~Drees, J.~R.~Ellis and C.~E.~M.~Wagner,
%``CPsuperH: A Computational tool for Higgs phenomenology in the minimal supersymmetric standard model with explicit CP violation,''
Comput. Phys. Commun. \textbf{156} (2004), 283-317
%doi:10.1016/S0010-4655(03)00463-6
[arXiv:hep-ph/0307377 [hep-ph]].
%300 citations counted in INSPIRE as of 04 Apr 2021

%\cite{Lee:2007gn}
\bibitem{Lee:2007gn}
J.~S.~Lee, M.~Carena, J.~Ellis, A.~Pilaftsis and C.~E.~M.~Wagner,
%``CPsuperH2.0: an Improved Computational Tool for Higgs Phenomenology in the MSSM with Explicit CP Violation,''
Comput. Phys. Commun. \textbf{180} (2009), 312-331
%doi:10.1016/j.cpc.2008.09.003
[arXiv:0712.2360 [hep-ph]].
%134 citations counted in INSPIRE as of 04 Apr 2021


%\cite{Cheung:2009fc}
\bibitem{Cheung:2009fc}
K.~Cheung, O.~C.~W.~Kong and J.~S.~Lee,
%``Electric and anomalous magnetic dipole moments of the muon in the MSSM,''
JHEP \textbf{06} (2009), 020
%doi:10.1088/1126-6708/2009/06/020
[arXiv:0904.4352 [hep-ph]].
%34 citations counted in INSPIRE as of 04 Apr 2021 

%\cite{Belanger:2006is}
\bibitem{Belanger:2006is}
G.~Belanger, F.~Boudjema, A.~Pukhov and A.~Semenov,
%``MicrOMEGAs 2.0: A Program to calculate the relic density of dark matter in a generic model,''
Comput. Phys. Commun. \textbf{176} (2007), 367-382
%doi:10.1016/j.cpc.2006.11.008
[arXiv:hep-ph/0607059 [hep-ph]].
%691 citations counted in INSPIRE as of 04 Apr 2021


%\cite{Evans:2020vil}
\bibitem{Evans:2020vil}
J.~Evans, C.~Han, T.~T.~Yanagida and N.~Yokozaki,
%``A Complete Solution to the Strong CP Problem: a SUSY Extension of the Nelson-Barr Model,''
[arXiv:2002.04204 [hep-ph]].
%2 citations counted in INSPIRE as of 07 Apr 2021





\end{thebibliography}
\end{document}